\documentstyle[12pt]{article}  
\newcommand{\be}{\begin{equation}}
\newcommand{\ee}{\end{equation}}
\newcommand{\ba}{\begin{eqnarray}}
\newcommand{\ea}{\end{eqnarray}}

\newcommand{\gsim}{\mathrel{\hbox{\rlap{\lower.55ex \hbox {$\sim$}}
                   \kern-.3em \raise.4ex \hbox{$>$}}}}
\newcommand{\lsim}{\mathrel{\hbox{\rlap{\lower.55ex \hbox {$\sim$}}
                   \kern-.3em \raise.4ex \hbox{$<$}}}}
\begin{document}
\begin{center}
{\bf The Blandford-Znajek Process}\\
{\bf as a}
{\bf Central Engine for a  Gamma Ray Burst}
\vskip 0.5cm
 Hyun Kyu Lee\footnote{corresponding author, 
E-mail: hklee@hepth.hanyang.ac.kr}  \\
Department of Physics,
Hanyang University, Seoul 133-791, Korea \\
R.A.M.J. Wijers and G.E. Brown \\

Department of Physics and Astronomy,
State University of New York at Stony Brook,
Stony Brook, NY 11794-3800
\end{center}
\vskip 0.5cm

\noindent {\bf Abstract:}
   
\noindent  
We investigate the possibility that gamma-ray bursts are powered by
a central engine consisting of a black hole with an external magnetic
field anchored in a surrounding disk or torus. The energy source is then
the rotation of the black hole, and it is extracted electromagnetically
via a Poynting flux, a mechanism first proposed by Blandford and Znajek
(1997) for AGN.

Our reanalysis of the strength of the Blandford-Znajek power shows that
the energy extraction rate of the black hole has been underestimated
by a factor ten in previous works. Accounting both for the maximum
rotation energy of the hole and for the efficiency of electromagnetic
extraction, we find that a maximum of 9\% of the rest mass of the hole can
be converted to a Poynting flow, i.e. the energy available to produce a
gamma-ray burst is $1.6\times 10^{53}(M/M_{\odot})$\,erg for a black hole of
mass $M$. We show that the black holes formed in a variety of gamma-ray
burst scenarios probably contain the required high angular momentum.

To extract the energy from a black hole in the required time of
$\lsim 1000$ \,s a field of $10^{15}$\,G near the black hole is needed. We
give an example of a disk-plus-field structure that both delivers the
required field and makes the Poynting flux from the hole dominate that of
the disk. Thereby we demonstrate that the Poynting energy extracted
need not be dominated by the disk, nor is limited to the binding energy
of the disk. This means that the Blandford-Znajek mechanism remains a
very good candidate for powering gamma-ray bursts.
       
\noindent  Key Words: gamma ray bursts, black hole, 
accretion disk 

\newpage
  
\section{Introduction}
Gamma ray bursts presently provide great excitement in astronomy and 
astrophysics as optical observations by way of many instruments give 
considerable detail of the history of each burst. We are concerned 
here with the prodigious energy in each burst, the estimate for GRB 971214 
being $\gsim 3 \times 10^{53} ergs$ \cite{kulkarni},
although this could be diminished if considerable beaming is involved in 
the central engine, as we will discuss.

Amazingly, $2 \times 10^{54} ergs$ is just the rest mass energy of our sun, 
so it seems immediately clear that the central engine for the GRB must be
 able to extract a substantial fraction of the rest mass energy of a compact
 object, neutron star or black hole, and convert it into energy of GRB.

The second criterion for the central engine is that it must be able to 
deliver power over a long interval up to $\sim 1000$ seconds, since some 
GRB's last that long although other GRB's last only a fraction of a second.
It must also be able to account for the vast diversity in pulses, etc., or,
alternatively, one must have a number of diverse mechanisms.

We believe the need to deliver power over the long time found in some bursters
to be the most difficult requirement to fulfill, since the final merger 
time of the compact objects is only a fraction of a second and it is difficult
 to produce a high energy source of, e.g., $\nu \bar{\nu}$-collisions that 
goes on for more than two or three seconds.

For many years mergers of binary neutron stars were considered to be likely 
sources for the GRB's. The estimated merger rate in our Galaxy of a few GEM
\footnote{ 1 GEM = 1 Galactic Event per Megayear} is of the right order for
the occurence of GRB's. The possible problem  with binary mergers might be 
the ejected materials during the merging processes.
Not more than $\sim 10^{-5}
M_{\odot}$ of baryons can be involved in the GRB, since it would not be 
possible to accelerate a higher mass of nucleons up to the Lorentz factors 
 $\Gamma \sim 100$  needed with the energies available. 

We find the merger of a neutron star with a black hole to be a particularly
attractive mechanism. The baryon number ``pollution" problem can be solved 
by the main part of the baryons going over the event horizon. In the
Blandford-Znajek mechanism\cite{BZ} we wish to invoke, a substantial proportion of
the rotational energy of the black hole, which will be sent into rapid
rotation by swallowing up the neutron star matter, can be extracted through 
the Poynting vector. The rate of extraction is proportional to the square of 
magnetic field strength, $B^2$, as we shall discuss, so that power can be 
furnished over varying times, depending upon the value of $B$. With 
substantial  beaming, we estimate that $B\sim 10^{15}$G would be sufficient
to power the most energetic GRB's with $\sim 10^{53}ergs$.

Recently at least three magnetars, neutron stars with fields 
$B\sim 10^{14}-10^{15}$G, have been observed.
 Their visible lifetime is only a few thousand years, 
because neutron stars slow down by emission of magnetic dipole radiation and
join the ``graveyard" where they no longer emit pulses. The time of 
observability is proportional to $B^{-1}$, so the number of these high 
magnetic field stars may not be an order of magnitude less than the 
garden variety 
$10^{12}$G neutron stars. It is also possible that existing magnetic fields 
can be increased by the dynamo effect. 

Failed supernovae were suggested by Woosley(1993)\cite{woosley}
as a source of GRB. In this case the black hole would be formed in the 
center of a massive star, and surrounding baryonic matter would accrete into 
it, spinning it up. This mechanism is often discussed under the title of 
hypernovae\cite{bp}. 

More recently Bethe and Brown(1998)\cite{BB} found that in binary 
neutron star evolutions, an order of magnitude more low-mass black-hole,
neutron-star binaries were formed than binary neutron stars. The low-mass 
black-hole mass of $\sim 2.4M_{\odot}$ looks favorable for the 
Blandford-Znajek mechanism.

In some calculations which begin with a neutron star binary, one of the 
neutron stars evolves into a black hole in the process of accretion, and the 
resulting binary might also be a good candidate for GRB's. In any case, there
 are various possibilities furnished by black-hole, neutron-star binaries.

In this paper we will  discuss the Blandford-Znajek mechanism in quantitative 
detail. In section 2, an  overview of the proposed central engine for 
gamma ray bursts using Blandford-Znajek process is given. The Blandford-Znajek
process and the evolution of the black hole will be discussed in detail 
in section 3 and section 4 respectively.  The structure of ambient magnetic 
field surrounding black hole and accretion disk is discussed in section 5. 
In section 6, we will give a rough estimation of the possible angular momentum 
of the black hole which might 
emerge as a final compact object during the merging or collapsing processes. 
The possible constraint from the surrounding accretion disk is discussed 
in section 7
and the results are summarized and discussed in section 8. 

\newpage

\section{Overview  of the Proposed Central Engine for GRB}

\noindent  Two decades ago Blandford and Znajek\cite{BZ}\cite{TPM} proposed a
process(BZ) in which rotational energy of black hole can be efficiently 
extracted.  
If there are sufficient charge distributions around the black hole 
to provide the force-free condition, then the magnetic field lines 
exert no force and corotate rigidly with  the rotating black hole.   The induced  current 
loops  loops which  
 pass  along the black 
hole's stretched horizon  feel the  forces by the  magnetic field supported 
by the environment. 
Hence these forces  give magnetic 
braking of the black hole rotation. 
The maximum amount of energy which can be extracted out of the 
black hole 
without violating the second  law of thermodynamics  is the rotational 
energy, $E_{rot}$, which is defined as 
\ba
E_{rot} &=& Mc^2 - M_{irr}c^2 \label{mrot}\\
\ea
where
\ba
M_{irr} &=& \sqrt{\frac{A_H c^4}{16 \pi G^2}}\label{mirr}\\
&=&  \sqrt{\frac{S_H}{4\pi k_B}} M_{planck} 
\ea
where $A_H$ and $S_H$ are surface area and entropy of a black hole 
respectively\footnote{For a solar mass black hole, $A_H \sim 10^{12} cm^2, 
S_H \sim 10^{77}k_B$. The Planck mass is
$M_{planck} = \sqrt{\frac{c \hbar }{G}} = 2.18 \times 10^{-5} g$.}
and $k_B$ is Boltzman's constant.
The rotational energy 
of a black hole with angular momentum $J$
 is a fraction of the black hole mass $M$,
\ba
E_{rot} &=& f(\tilde{a}) Mc^2\\
f(\tilde{a}) &=& 1- \sqrt{\frac{1}{2}[1+\sqrt{1-\tilde{a}^2}]}
\ea
where $\tilde{a}= \frac{Jc}{M^2G}$ is the rotation parameter.
For a maximally rotating black hole ($\tilde{a}=1$), 
$f=0.29$. 
 In BZ, the efficiency of extracting the rotational energy is determined by 
the ratio between the angular velocities of black hole, $\Omega_H$, and 
magnetic field angular velocity $\Omega_F$,
\ba
\epsilon_{\Omega} = \frac{\Omega_F}{\Omega_H} \label{epsilon}
\ea     
The rest of the rotational energy is dissipated into the black 
hole 
increasing the 
entropy or equivalently irreducible mass\footnote{The mass of a nonrotating 
black
hole itself is its irreducible mass}. 
The total BZ    energy  available is
\ba
E_{BZ} = 1.8 \times 10^{54} \epsilon_{\Omega} f(\tilde{a}) \ \
(\frac{M}{M_{\odot}})
 erg
\label{ebz}
\ea  
For the optimal processes 
$\epsilon_{\Omega} \sim 0.5$
\cite{TPM}.   

Since the energy transport is in the form of Poynting flow 
in BZ the 
outgoing energy flux from the black hole is basically $B^2c$. 
An order of magnitude calculation is now in order.      
 There are basically three parameters: Mass of the black hole $M$
 and magnetic field on the horizon $B$, which are dimensionful,  and  angular 
momentum parameter
of the black hole $\tilde{a}$.  The time scale for the BZ process can be 
calculated by 
the ratio  of  the black hole mass to  the output power from the black 
hole surface 
$B^2R^2c$
\ba
\tau_{BZ} &\sim& \frac{Mc^2}{B^2 R^2c} \sim \frac{Mc^5}{B^2 M^2G^2}\nonumber\\
          &=& \frac{c^5}{B^2 M G^2} \ \ = \ \ 2.7 \times 10^3 
(\frac{10^{15}gauss}
          {B})^2 (\frac{ 
          M_{\odot}}{M}) s \label{tbz}
\ea
where  
$ M_{\odot}^{-1} (10^{15}gauss)^{-2}\frac{c^5}{G^2} = 2.7 \times 10^3 s$ 
and the radius of horizon, $R$,
 is taken to be $\sim \frac{GM}{c^2}$.  Also the outgoing 
Poynting power is
\ba
P_{BZ} \sim B^2 R^2c = 6.7 \times 10^{50}  (\frac{B}{10^{15} gauss})^2 
(\frac{M}{M_{\odot}})^2 erg/s.\label{power0}
\ea
\noindent The fluence of the recently observed GRB971214\cite{kulkarni} corresponds 
to 
$E_{\gamma} = 
10^{53.5}(\frac{\Omega_{\gamma}}{4\pi}) erg$ which is consistent with $E_{BZ}$.
That of  GRB990123 may be as large as $E_{\gamma} =
3.4 \times
10^{54}(\frac{\Omega_{\gamma}}{4\pi})$\cite{batse}. This suggests 
that if a strong enough magnetic field ($\sim 10^{15}$) 
on the black 
hole can be supported by the surrounding material(accretion disk) the BZ
process is  a good candidate to provide the powerful energy of the 
GRB in the 
observed time interval up to 1000s , which is comparable to the BZ time scale 
$\tau_{BZ}$.   

In recent years a black hole plus debris torus system(or accretion disk) has 
been 
considered to be a  plausible structure for the GRB central engine\cite{ralph}.
 The 
presence of the accretion disk is important for the BZ process  because it is 
the supporting system of the strong
magnetic field on the black hole, which would disperse without the pressure 
from the fields anchored in the accretion disk. 
Recent numerical calculations\cite{popham} show 
that accretion disks formed by various  merging processes are found to have 
large  enough
pressure such that they can support $\sim 10^{15}G$ assuming a value of the 
disk viscosity parameter $\alpha
\sim 0.1$, where $\alpha$ is the usual parameter in  scaling the viscosity. This gives a 
relevant  order of magnitude  magnetic field on the 
black 
hole,  which is, however,  not
considered to be much larger than magnetic field of the inner accretion disk
\cite{lop}\cite{ghosh}.  The discovery that soft gamma ray 
bursts are magnetars
\cite{sgr} 
also supports the presence of the strong magnetic fields of  
$\sim 10^{15}G$ in nature.  
The identification by now of
three soft gamma repeaters as strong-field pulsars indicates that there may be
a large population of such objects:
since the pulsar spindown times scale as $B^{-1}$,
we would expect to observe only 1 magnetar for every 1000 normal pulsars if they
were formed at the same rate, and if selection effects were the same for the two
populations. We see 3 magnetars and about 700 normal pulsars, but since they
are found in very different ways the selection effects are hard to quantify.
It is nonetheless clear that magnetars may be formed in our Galaxy at a rate
not very different from that of normal pulsars.

\noindent The life time of the accretion disk is also very important 
for the GRB time 
scale because it supports the magnetic field on the black hole. According to 
numerical simulations of 
merging systems which evolve eventually into black hole - accretion disk 
configuration\cite{popham} the viscous life times are 0.1 - 150 s, which are 
not inconsistent with the GRB time scale.  Also it has been  pointed 
out\cite{ralph} that 
a residual cold disk of $\sim 10^{-3} M_{\odot}$ can support $10^{15}G$, even 
after the major part of the accretion disk has been drained into the black hole
 or dispersed 
away.  

Recent hydrodynamic simulations of merging neutron stars and black 
holes\cite{jre} show that along the rotation axis of the black hole an almost 
baryon-free funnel is possible. This can be easily understood since the 
material 
above the hole axis has not much angular momentum so that it can be drained 
quickly, leaving a baryon-free funnel.  
Hence relativistically expanding 
jets along the funnel, fueled by Poynting outflow which is  collimated along the
 rotation axis\cite{fendt},  can give rise to gamma ray 
bursts effectively.   
\noindent It has been observed that the BZ process is also possible from the 
disk since the  magnetic field on the disk is not much less than that on the 
black hole\cite{lop}.  However the energy outflow from the disk is mostly 
directed vertically from  the disk where 
the baryon loading is 
supposed to be relatively high enough to keep the baryons 
 from being highly relativistic.  
Therefore the BZ process from  the disk 
can be considered to have  not much to do with gamma ray 
burst phenomena. However the BZ from the disk could power an outflow with lower 
$\Gamma$, but nonetheless high energy, which could cause an afterglow 
at large angles. That would lead to more afterglows being visible than GRBs, 
because the afterglows are less beamed.
The BZ mechanism can also play a very important role in the disk accretion
because it carries out angular momentum from the disk.
We will discuss this in more detail later.    
           
The structure we are proposing as a central engine of the GRB is a system of 
black hole
 - accretion disk(debris of the torus): 

\noindent The rotating black hole is threaded by a strong 
magnetic field. Along the baryon-free funnel relativistic jets fueled by 
Poynting outflow give rise to the GRB.  The interaction between disk and black 
hole is characterized by accretion  and magnetic coupling. 
We consider  the BZ process only
after the main accretion process is completed, leaving an accretion disk of
cold  residual material, which can support a strong enough magnetic field.

\section{Blandford-Znajek Process}

Consider a half hemisphere(radius $R$) rotating with angular velocity $\Omega$  
and a  circle on the surface at fixed $\theta$( in the spherical polar
 coordinate 
system) across which a surface current $I$ flows down from the pole. When the 
external magnetic field $B$ is imposed to thread the surface outward normally,
the surface current  feels a force and  the  
torque due to the Lorentz 
force exerted by the annular ring of  width $Rd\theta$ is
\ba
\Delta T &=& - R \sin \theta \times  I R \Delta B \theta \\
  &=& - \frac{I}{2\pi} B \Delta A_{ann.} \\
  &=& - \frac{I}{2\pi} \Delta \Psi\label{torque} 
\ea
where $\Delta \Psi$ is the magnetic flux through annular ring extended by 
$d\theta$ with surface area $A_{ann}$. We consider an axially symmetric 
situation.
 From this magnetic braking, we can calculate the rotational energy loss rate
\ba
\Delta P_{rot} &=& \Omega \times \Delta T \\
        &=& \Omega \frac{I}{2\pi} \Delta \Psi.\label{dprot}
\ea

Blandford and Znajek\cite{BZ} demonstrated that such a  magnetic braking 
is possible, provided that the external charge distribution can support  
an electric current with the magnetic field threading the horizon.  
The original formulation of Bland ford and Zanjek\cite{BZ} is 
summerized in Appendix 3. Macdonald and 
Thorne\cite{TM} reformulated the Blandford-Znajek process using a 
3+1 dimensional formalism,
 in 
which the complicated physics beyond the horizon can be expressed in terms 
of 
physical quantities defined on the stretched horizon\cite{TPM}.  It can be shown that
the rotational energy  loss due to magnetic braking can be obtained from 
eq.(\ref{dprot}) by simply replacing $\Omega$ and $ B$ with
\ba
\Omega \rightarrow \Omega_H , \ \  B \rightarrow B_H
\ea
where $H$ denotes the quantity on the stretched horizon \cite{TPM}:
\ba
\Delta P_{rot} &=& \Omega_H \times \Delta T \\
        &=& \Omega_H \frac{I}{2\pi} \Delta \Psi.\label{dproth}
\ea
The current $I$ induced 
by the black hole rotation and 
 the angular velocity  $\Omega_F(\theta)$ of the rigidly
rotating magnetic field 
which is dragged by the rotating black hole can be determined together with
the magnetic field  
by solving  Maxwell's 
equations:
\ba
F^{\mu\nu}_{\,\,\, ;\nu} = 4 \pi J^{\mu}\label{maxs}
\ea
with the force free-condition 
\ba
F^{\mu\nu}J_{\nu} = 0,\label{ffree}
\ea
where $J^{\mu}$ is a current density vector.
The detailed structure of the magnetic field will be discussed 
in section 5. 
\noindent In the BZ process, however, the power which can be carried out as
Poynting outflow along the field lines is 
\ba
\Delta P_{mag} = \Omega_F \frac{I}{2\pi} \Delta \Psi  
\label{protm}
\ea
Then the rest of the rotational energy
is used to
increase the entropy(or equivalently irreducible mass) of the black hole.
Therefore the efficiency of the BZ process which is defined as the ratio of 
$P_{mag}$ to $P_{rot}$ is
\ba
\epsilon_{\Omega} = P_{mag}/P_{rot} = \frac{\Omega_F}{\Omega_H} \label{effi}
\ea 
The ideal efficiency, $\epsilon_{\Omega}=1$, is meaningless because 
for $\Omega_{F}
=\Omega_H$ the Poynting outflow itself is zero as can be seen below in 
eq.(\ref{i}).  The optimal power can be obtained at
 $\epsilon_{\Omega} = 1/2$\cite{TM}.  Then the rest of the rotational energy 
is used to 
increase the entropy(or equivalently irreducible mass) of the black hole.
Now consider the loading region far from the black hole, onto which magnetic
fields out of the black hole anchor.  In most of the cases we are interested 
 in, the inertia of  the 
loading region  can be considered to be so large that the transported 
angular momentum cannot give rise to any substantial increase of the  angular 
velocity of the loading region.  Therefore   
the angular velocity of the loading region can be assumed to be zero and 
the 
power delivered by the torque, eq.(\ref{torque}), 
along the field line is given 
\ba
\Delta P_{L} 
&=& \Omega_F \Delta T \label{pl0}\\
  &=& \Omega_F \frac{I}{2\pi} \Delta \Psi\label{dpl}
\ea
which is identified as the BZ power for GRB:
\ba
\Delta P_{BZ} = \Delta P_{mag} = \Delta P_{L}
\ea
Since the current $I$ induced
by black hole rotation 
is given by\cite{TPM}
\ba
I(\theta) = \frac{1}{2c} (\Omega_H - \Omega_F(\theta))
\tilde{\omega}^2 B_H\label{i}
\ea
 the rotational energy loss and the BZ power can be given by 
\ba
\Delta P_{rot} &=& \frac{\Omega_H (\Omega_H - \Omega_F)}{4\pi}
\tilde{\omega}^2 B_H\Delta \Psi \label{proti}\\
\Delta P_{BZ} &=& \frac{\Omega_F (\Omega_H - \Omega_F)}{4\pi c}
\tilde{\omega}^2 B_H \Delta \Psi, \label{pl}
\ea    
where  $\tilde{\omega}$ is a kind of
cylindrical radius defined in Boyer-Lindquist coordinates
(Appendix 1). 
 
 To get the total power, eq.(\ref{pl}) 
should be integrated 
from $\theta \sim 0$   to $\theta_{BZ}$, up to which the magnetic 
field lines from the black hole anchored on to the loading region 
for GRB. 
As a first approximation, we put
\ba
\theta_{BZ} &\sim& \pi/2\\
B_H(\theta) &\sim& <B_H>
\ea
and we get an optimal power
\ba
P_{BZ} = 1.7\times 10^{50}  \tilde{a}^2 (\frac{M}{M_{\odot}})^2
    (\frac{<B_H>}{10^{15}gauss})^2 f(h) \, erg/s. \label{powerm}
\ea 
The details are given in Appendix 2.
The rest of the rotational energy given by
\ba
P_{H} &=& P_{rot} - P_{BZ}\\
      &=& \frac{\Omega_H-\Omega_F}{\Omega_F} P_{BZ}\\
      &=& P_{BZ}, \ \ for \ \ optimal  \ \ case
\ea
is dissipated into the black hole increasing the irreducible part of the 
black hole
mass(equivalently increasing the entropy of the black hole).
 
For $\theta_{BZ} <  \ \ \theta \ \ \leq \pi/2$, the magnetic field lines 
from the black hole can be anchored onto elsewhere than the loading region,
for example, onto the inner accretion disk.  Then the angular velocity of
the field line is determined by the angular velocity of the disk  
$\Omega_D$, $\Omega_F = \Omega_D$.
The innermost
radius of the disk can be considered to be the marginally stable orbit,
$r_{ms}$ \cite{st}. For an orbit rotating in the same direction as the  
black hole 
rotation, we have:
\ba
r_{ms} &=& \frac{G}{c^2}M[3+Z_2 - \sqrt{(3-Z_1)(3+Z_1 + 2Z_2)}],\label{rms}\\
  Z_1 &=& 1+(1-\tilde{a}^2)^{1/3}[(1+\tilde{a})^{1/3} +(1-\tilde{a})^{1/3}],\\
  Z_2 &=& \sqrt{3\tilde{a}^2 + Z_1^2}.
\ea
Here $r_{ms}$ becomes $GM/c^2$ as $\tilde{a} \rightarrow 1$ (extreme rotation). 
In this limit, 
the angular velocity of disk $\Omega_D \sim \Omega_H$. Then one can expect 
$\Omega_F \sim \Omega_H$ so that there is no Poynting flow to the disk. 
However in this case there is no BZ Poynting outflow.  
For finite $\tilde{a} < 1$,  we can have either  $
\Omega_H \ \  > \ \ \Omega_D$ or $\Omega_H \ \ < \ \ \Omega_D$, which
can be determined by solving eq.(\ref{maxs}). 
However if we can assume that 
the power 
from/to the 
disk can be much suppressed compared to that of the  loading region, 
we can ignore the magnetic coupling between the black hole and the 
accretion disk.  For example, suppose  
the  portion of black hole - threading magnetic fields  
anchored onto the 
disk is somehow suppressed, $ \theta_{BZ} \sim  \pi / 2$, then we can 
assume   
 there is no significant 
energy  and 
angular momentum feedback into the disk due to magnetic coupling. If it 
is not so we can 
discuss only the limiting case.  
 
\section{Evolution of a black hole via the Blandford-Znajek process}

While the black hole is slowed down and a part of the rotational energy is 
carried
 out as Poynting outflow, the rest of the rotational energy increases the 
entropy 
of the black hole or its irreducible mass. The increasing rate of the 
irreducible mass 
 is  given by using eq.(\ref{proti}) and (\ref{pl})
\ba
\frac{d M_{irr}}{dt} &=&  P_{rot} -  P_{L} \\
   &=& \int \frac{(\Omega_H - \Omega_F)^2}{4\pi c}
\tilde{\omega}^2 B_H \Delta \Psi. \label{dmirr}
\ea
The irreducible mass eventually becomes the mass of Schwarzshild black 
hole when 
it stops rotating. The  difference between the initial Kerr black hole mass,
 $M_o$, and the final Schwarzshild mass, $M$, is the energy output from black 
hole via the Blandford-Znajek process. 
The evolution of a Kerr black hole is determined by the evolution of 
its mass and angular momentum given by 
\ba
\frac{d M c^2}{dt} &=& -P_{L} \\
\frac{dJ}{dt} &=&  T\label{jt}
\ea
Using eq.(\ref{pl0}) and (\ref{jt}) in the Blandford-Znajek process, 
the evolution 
of the mass 
and the angular momentum are related  by
\ba 
\frac{d M}{dt} = \Omega_F \frac{dJ}{dt}\label{mj} 
\ea
For the optimal case, $\epsilon_{\Omega} = 1/2(\Omega_F=\Omega_H/2)$, we 
can obtain an analytic 
expression for the mass in terms of the angular momentum. With the angular 
velocity of 
a black hole expressed in the angular momentum given by in the natural units
$G=c=1$ 
\ba
\Omega_H = \frac{J}{2M^3(1+\sqrt{1-J^2/M^4})}
\ea
we get 
\ba
\frac{dM}{dt} 
&=& \frac{J}{4M^3(1+\sqrt{1-J^2/M^4})}
\frac{dJ}{dt}\\
2\frac{dM^4}{dt} &=& \frac{1}{1+\sqrt{1-J^2/M^4}}
\frac{dJ^2}{dt}. \label{j2t} 
\ea
Intoducing a new variable $H$ defined as
\ba
H = \frac{r_H}{M} = 1 + \sqrt{1-\frac{J^2}{M^4}},\label{hdef}
\ea
eq.(\ref{j2t}) can be written   
\ba
2\frac{dM^4}{dt} &=& \frac{1}{H}
\frac{dJ^2}{dt}.\label{Hmj}
\ea
Using the identities
\ba
\frac{d}{dt}(\frac{J^2}{M^4}) &=& 2(1-H)\frac{dH}{dt}\\
\frac{dJ^2}{dt} 
&=& \frac{J^2}{M^4}\frac{dM^4}{dt} + M^4 2(1-H)\frac{dH}{dt}\\
J^2/M^4 &=& 2H - H^2\label{HJ}
\ea
eq.(\ref{Hmj}) can be written
\ba
\frac{1}{M} \frac{dM}{dt} &=& \frac{(1-H)}{2H^2}\frac{dH}{dt},\label{mh}
\ea
which can be integrated analytically to give\cite{ok}   
\ba
M = M_o e^{\frac{H-H_o}{2HH_o}}\sqrt{\frac{H_o}{H}},\label{mhf}
\ea
where $H_o$  and $M_o$ represent the initial angular momentum and mass of the
black hole. 
 From eq.(\ref{hdef}) one can see that $H = 1$ for the maximally 
rotating ($\tilde{a} =1$) black hole and 
$H=2$ for the 
nonrotating one.  

\noindent Consider the black hole initially maximally rotating,
which is slowed down by the Blandford-Znajek mechanism in the optimal 
mode($\Omega_F = \Omega_H/2$)\footnote{The more general discussion has been
 given by I. Okamoto\cite{ok} where $\zeta = 1/\epsilon_{\Omega} -1$.}.  
The final black hole mass is  given by
\ba
M &=& \frac{e^{1/4}}{\sqrt{2}} M_o\\
  &=& 0.91 M_o. \label{0.91}
\ea
We see that 9 \% of the initial mass or 31 \% of the rotational 
energy can be taken 
out to power the gamma ray burst from  the maximally rotating black hole. The 
extracted energy is therefore less than a half of the initial 
rotational energy. 
This can be easily understood by noticing that the fraction of the rotational 
energy
drops faster because of the decreasing total mass and at the same time 
increasing irreducible mass in eq.(1). For 
$\tilde{a} = 0.5$, $M= 0.98M_o$ or 2 \% of the initial mass 
can be used to power 
the gamma ray burst.       
The time dependence of the power can be obtained from eq.(\ref{power}) using 
eq.(\ref{HJ}):
\ba
P &=& -\frac{dM c^2}{dt} = \frac{f(h)}{4} (2H-H^2)M^2 B_H^2 
\frac{G^2}{c^3}, \label{dmdth} 
\ea
where $f(h)$ is defined in Appendix 2 and $H = \frac{2}{1+h^2}$.  
The initial rate for the maximally 
rotating black hole can be obtained by taking $ H = 1$ and $ f = \pi -2$. 
Eq.(\ref{dmdth}) can be written as
\ba
\frac{1}{M^2}\frac{dM}{dt} = -\frac{\pi -2}{4} B_H^2\frac{G^2}{c^5}.
\ea
Assuming there is no change in magnetic field which is supported by the 
environment,  we get
\ba
M = \frac{M_o}{1+ (\pi -2)M_o B_H^2G^2 t/4c^5}.\label{mt1}
\ea
If we  extrapolate eq.(\ref{mt1}) until the black hole stops rotating, 
$M \rightarrow  0.91 \, M_o$, we get the time scale of Blandford-Znajek 
process as 
\ba
\tau &=&
\frac{0.35 c^5}{M_oB_H^2 G^2}\\
     & \sim & 10^3 (\frac{10^{15}gauss}
          {B_H})^2 (\frac{
          M_{\odot}}{M}) s. \label{tbz1}
\ea
Since there is no considerable change of the weighting factor $f(h)$ in 
eq.(\ref{dmdth}) from the maximally rotating black hole  to the nonrotating 
black hole($f = 2/3$), Eq.(\ref{tbz1}) can be considered as a reasonable 
estimate of a time scale, which is consistent with the rough estimate given 
in eq.(\ref{tbz}).   

\section{Magnetic field and Force-free plasma}

The rotating black hole immersed in the magnetic field induces an electric 
field 
around the black hole\cite{TPM}. 
\noindent  The electromagnetic field 
in the vicinity of rotating black hole immersed in the uniform magnetic field 
in free space  has been obtained 
\cite{TPM}\cite{wald}\cite{king} as  a solution of the source-free Maxwell
equation
\ba
F^{\mu\nu}_{\,\,\, ;\nu} = 0.\label{maxf}
\ea
 From the analytic expression(Appendix 2) which gives an  asymptotically uniform
magnetic field at infinity($r \rightarrow \infty$)
\ba
\vec{B} = B \hat{z}, \,\,\, r \rightarrow \infty, \label{binfty}
\ea
one can see that  the radial component of the  magnetic field 
on the horizon, $B_r^H$, 
\ba
B^H_r &=& \frac{B\cos\theta}{(r_H^2 +(a/c)^2 \cos^2 \theta)^2}  
  [(r_H^2 -(a/c)^2)
(r_H^2 -(a/c)^2 \cos^2 \theta) \nonumber \\
  & & \, \, \, + 2(a/c)^2 r_H (r_H-M)
(1 + \cos^2 \theta)]\label{bhr}\\
\ea
vanishes as the rotation of black hole approaches extreme
rotation
\ba
a \rightarrow GM/c , \,\,\,\,\, 
r_H \rightarrow GM/c^2.
\ea
This is what has been observed \cite{king}\cite{bicak} as the absence of
the magnetic
flux across the maximally rotating black hole.
One can also see that there is no outward Pointying flow, which 
means no outflow  of energy to infinity.
Essentially it is because of the abscence of the toroidal component of
the magnetic field due to the vacuum environment which is 
charge and current free space.
In other words, there is no current on the stretched horizon on 
which the external 
magnetic field exerts torques to slow the black hole 
down  so as to extract its rotational energy.  
Therefore it is neccessary to have a
magnetosphere with charges and currents to extract 
the rotational energy of the black hole. The force-free magnetosphere around 
a rotating black hole has been
 proposed \cite{BZ}:
\ba
F^{\mu\nu}_{\,\,\, ;\nu} = 4 \pi J^{\mu}\label{maxs}
\ea
where $J^{\mu}$ is a current density vector and
\ba
F^{\mu\nu}J_{\nu} = 0\label{ffree}
\ea
which is the force-free condition. From eq.(\ref{ffree}), we get the degenerate 
condition
\ba
\vec{E} \cdot \vec{B} = 0 \label{deg}
\ea
In the case of a rotating black hole   in charge-free space 
one can see 
 using eqs. in Appendix 2  
\ba
\vec{E}^{H} \cdot \vec{B}^H   \neq 0\label{ebh}
\ea
It should be remarked that since the force-free condition is essential 
for the BZ 
process the expulsion of magnetic field on  the rotating black hole demonstrated for the 
charge-free space cannot be directly addressed in the 
 BZ process\cite{bicak}. 

\noindent To maintain the current flows charged particles
(electrons and positrons)
should be supplied by a source outside the horizon since the black 
hole itself
cannot provide the outgoing particle from inside the horizon.  Blanford and 
Znajek\cite{BZ} proposed that the  strong 
electric field induced by the rotating black hole  can give  rise to a spark gap in which 
sufficient charged particles are created to supply the 
currents\cite{hok}.  Another mechanism provides  charged 
particles around black hole: electron-positron pair creation by neutrino 
annihilation\cite{wb}.  Recent numerical studies of merging binary systems
\cite{popham}\cite{RJ} which result in black hole-accretion disk 
configurations show 
that the power of electron-positron pairs by the annihilation of 
neutrinos and antineutrinos which are radiated out of the disk is  
\ba
\dot{E}_{\nu  \bar{\nu}} \sim 10^{50} erg/s.
\ea
This power is being poured into  the space above the black  hole for $0.01 - 1  s$.  If 
we
divide it by the average neutrino energy $<\epsilon_{\nu}> \sim 10 MeV$ 
\cite{RJ}, we get a rough
estimate of the numbers of $e^+   e^-$ pairs
\ba
\dot{N}_{pair} \sim 10^{56}/s
\ea
or the charge producing rate for $e^+$(or $e^-$;the sum is always zero) is
\ba
\dot{Q}_e =  e \dot{N}_{pair} \sim 10^{37} C/s\label{qdot}
\ea
The magnitude of the currents involved in the optimal BZ process can be 
obtained from eq.(\ref{powerm}) 
\ba
I \sim   \sqrt{\frac{10^{50} erg/s}{R_H}} \sim 10^{20} C/s\label{I}
\ea
for a black hole with solar mass threaded by a $10^{15}G$ magnetic field. $R_H$ is the 
surface resistance of the horizon, 377 Ohm. From the comparison of 
eq.(\ref{qdot}) with   (\ref{I}), the   neutrino annhilation   process produces   orders of 
magnitude 
more than enough pairs to keep the necessary currents for the optimal 
BZ process. 
The possible effects of neutrino annihilation during the active neutrino 
cooling of the
accretion disk is that  the magnetosphere for the BZ process might be disturbed so
much that the BZ process is suspended until the burst of $e^+ \ e^-$ pairs  
clears out.  
However, since the pairs  are produced with strong directionality along the 
rotation 
axis,  most of the produced pairs expand  along  the axis in less than
 a second\cite{RJ}. This is reasonable because the  electric field along the magnetic 
field lines  is almost negligible and therefore it will take too long for  
the $e^{+}$ to 
reverse its velocity to make the same current as the $e^-$. Also the strong 
magnetic field keeps them from moving  perpendicular  to the 
magnetic field lines so that the separations between them cannot be effective. 
Therefore the pair contribution to the net current flow can be  
 negligible and it might 
not disturb the magnetosphere for the BZ process so violently
\footnote{Although very hard to estimate, a very small fraction of pairs which 
has very small momentum component  along the axis can contribute to BZ currents
along the magnetic field lines, which might be sufficient for the force-free 
configuration.}.  However,
the effects of the neutrino cooling process of the accretion disk can be 
considered as 
an additional disturbing burst by the $\nu\bar{\nu}$  driven  $e^+e^-
$ burst, which lasts less than few seconds of the  BZ  burst.
  
The structure of a force-free magnetosphere can be  described by a  
stream function $\psi(r, \theta)$ and two functions of $\psi$; $\Omega_F(\psi)$ and 
$B_{\phi}(\psi)$
(or equivalently $I= -\frac{1}{2}\alpha \tilde{\omega} B_{\phi}$)
\cite{BZ}\cite{md}. 
Hence $\psi$ at a point of 
$(r, \theta)$ is equal to the total magnetic flux upward through the azimuthal 
loop at $(r, \theta)$.  On the other hand from the equation of motion, 
the stream function is proportional to 
the toroidal component of the vector potential, $\psi = 2\pi A_{\phi}$.  
On the horizon, it determines the total magnetic flux, $\Psi(\theta)$, through
 the horizon up to  $\theta$,  $\Psi(\theta) = \psi(r_H, \theta)$.  
The poloidal and toroidal components of the electromagnetic field are
\ba
\vec{E}^P &=& - \frac{\Omega_F -\omega}{2\pi \alpha} \vec{\nabla}\psi, \ \ 
\vec{E}^T = 0 \\   
\vec{B}^P &=& \frac{\vec{\nabla}\psi \times \hat{\phi}}{2\pi \tilde{\omega}}, 
\vec{B}^T = -\frac{2I}{\alpha \tilde{\omega}} \hat{\phi}
\ea
where $I(r, \theta)$ is the total current  downward  through the loop at $(r, 
\theta)$. $\Omega_F(\psi)$ is the angular velocity of the magnetic field 
relative to absolute space. 
\noindent Blandford and Znajek\cite{BZ} derived the  differential equation
(stream equation) for the stream function, which  
takes the form\cite{md}
\ba
\vec{\nabla} \{ \frac{\alpha}{\tilde{\omega}^2}[1-\frac{(\Omega_F^2 -\omega^2) 
\tilde{\omega}^2}{\alpha^2}]\vec{\nabla}\psi \} + \frac{(\Omega_F - \omega)}
{\alpha}\frac{d\Omega_F}{d\psi}(\vec{\nabla}\psi)^2 + \frac{16\pi^2}{\alpha 
\tilde{\omega}^2}I \frac{dI}{d\psi} =0.\label{streameq}
\ea
It includes two functions of $\psi$, which are not known {\it a priori}. In 
solving eq.(\ref{streameq}) the boundaries of the force-free region and 
the boundary 
conditions on $\psi, \Omega_F,$ and $I$ should be specified. There is 
 no known analytical solution of eq.(\ref{streameq}).  Blandford and Znajek 
obtained perturbative solutions 
for small $a/M$ using the analytic solutions in charge free space around 
the non-rotating Schwarzschild black hole.  With $\Omega_F=0, \omega =0$ and 
$I=0$,
the stream equation reduces to 
\ba
\vec{\nabla} \{ \frac{\alpha}
{\tilde{\omega}^2}\vec{\nabla} \psi \} =0.\label{streamfree}
\ea
MacDonald\cite{md} developed a numerical method to obtain  solutions with
 finite $a/M$, in which the  solutions of eq.(\ref{streamfree}) with 
appropriate 
boundary conditions are spun up 
numerically.  It has been found that the poloidal field structure does not 
change greatly as the holes and fields are spun up.  This result implies that 
the flux of the poloidal magnetic field 
threading black hole does not decrease greatly  as $a \rightarrow M$ in 
contrast to the 
 rotating black hole in free space.

\noindent  One of the interesting solutions is the poloidal field structure
generated by the  paraboloidal magnetic field solution of eq.(\ref{streamfree}):\ba
\psi = \frac{\psi_0}{4 \ln 2} \{(r-2)(1-\cos \theta) + 2[2 \ln 2 - 
(1+\cos \theta) \ln (1 + \cos \theta)] \} \label{paramd}
\ea
where $\psi_0$ is the total flux threading the hole. Since this solution 
extends not only onto the horizon but also onto the equatorial plane 
where the accretion 
disk is 
supposed to be placed, the boundary conditions on the disk should be 
satisfied by the spun up solutions generated by eq.(\ref{paramd}).  
The boundary conditions depend strongly on the accretion disk model, which will
 be discussed in the next section.  The  presence of the accretion disk  may be 
considered to be the main source of the difficulty in obtaining a solution 
with
these complicated boundary conditions.  However, the solutions 
with proper boundary conditions provide us a way as to how we can infer the 
magnetic 
field from those  developed in the disk.  
It has been demonstrated\cite{lop}\cite{ghosh}  that
 the strength of the magnetic field on the black hole is not much stronger 
than those on the inner disk. 

\section{Rotation of the Black Hole}
It has been suggested that  merging compact binary systems\cite{ralph} 
or hypernovae
\cite{bp} may result
in a disk with rapidly rotating  black hole at the center.  In this section 
we will 
demonstrate how this is possible using semiquantitative arguments.  The basic 
idea is that a substantial part of the  orbital angular momentum 
of the binary system or the spin angular momentum of the progenitor of 
hypernova
can be imparted onto the black hole.

Consider first the BH-NS merger.  The typical distance for the merging system 
in 
this 
case is the tidal radius, $R_t$, at which the neutron star(radius $R_{NS}$) 
fills the Roche lobe\cite{kopal}:
\ba
R_t = \frac{R_{NS}}{0.46}(\frac{1+q}{q})^{1/3},\label{rt}    
\ea
where
\ba
q &=& \frac{M_{NS}}{M_{BH}}
\ea
If the tidal radius is greater than the last stable orbit radius
\footnote{This leads to the condition for the black hole mass\cite{BC}:
$M_{BH} < M_{NS}[(0.4\frac{R_{NS}c^2}{GM_{NS}})^{3/2} -1] \sim 2.3M_{\odot}$.}, 
$R_{ls}$, we can calculate the Keplerian orbital angular velocity 
\ba
\Omega_K = \sqrt{\frac{GM}{R_t^3}},
\ea
where 
\ba
M = M_{BH} + M_{NS}.
\ea
Then the orbital angular momentum of the binary system can be written 
\ba
J_{binary} &=& \mu R_t^2 \Omega_K \\
  &=& M_{BH} M_{NS} \sqrt{\frac{GR_t}{M}}
\ea
where $\mu$ is the reduced mass, $\mu = \frac{M_{BH}M_{NS}}{M}$.
Using eq.(\ref{rt}),
\ba
J_{binary} &=& 1.47 \  M_{BH} M_{NS} \sqrt{\frac{GR_{NS}}{M}}
(\frac{M}{M_{NS}})^{1/6}
\ea
During the collapse, a part of angular momentum is carried off  by 
gravitational waves(or possibly in the later stage by neutrino cooling ) and 
a part of the total mass explodes away or remains in the torus around the 
black hole. Assuming that a fraction($xy$) of the orbital angular momentum 
goes into 
the black 
hole, which keeps  a fraction($y$) of the total mass 
$M$\footnote{$x$ is a fraction of the specific angular momentum}, 
we can calculate 
the angular 
momentum parameter of the black hole, $\tilde{a}$:
\ba
J_{BH} &=& \tilde{a} (yM)^2\frac{G}{c} = xy J_{binary} \\
\tilde{a} &=& 1.47 \frac{x}{y} \frac{M_{BH} M_{NS}}{M^2} 
\sqrt{\frac{R_{NS}c^2}{GM}}(\frac{M}{M_{NS}})^{1/6}\label{abh}
\ea
For $M_{BH}=2.5M_{\odot}, M_{NS}= 1.5M_{\odot}, R_t = 10^6 cm$,
\ba
\tilde{a} = 0.53\frac{x}{y}
\ea
which is a quite reasonable value for an efficient Blandford-Znajek process.

For the NS-NS merger, the radius of physical contact(2$R_{NS}$) is smaller 
than the
tidal radius,
\ba
R_t = 2.46 R_{NS} 2^{1/3} = 3 R_{NS}.
\ea
Hence it is reasonable to consider the orbital angular momentum at the 
tidal radius. 
Following the same procedure as in the BH-NS merger, we get 
\ba
\tilde{a} &=& 0.31\frac{x}{y}\sqrt{\frac{R_{NS}c^2}{GM_{NS}}}\nonumber\\
     &=& 0.67\frac{x}{y}
\ea  
where we replaced $M_{BH}$ by $M_{NS}$ in eq.(\ref{abh}).

In the hypernova model\cite{bp}\cite{iwamoto}, a massive rapidly spinning 
progenitor
($M_o \sim 40M_{\odot}$)  is supposed to  
collapse into a rapidly  rotating black hole.  
If we assume a critically   rotating progenitor($\Omega = \Omega_K$ at the 
surface, $R_o$) and solid body rotation throughout,  
the angular momentum of the core which 
eventually  collapses into 
the black hole is
\ba
J_{core} &=&\frac{2}{5} M_{core} R_{core}^2 \Omega_K(R_o)
\ea
where the moment of inertia of the core is assumed be that of a uniformly 
distributed 
spherical object\footnote{Since in general 
$I=k^2M_{core}R_{core}^2$, and $k^2 << 1$ for radiative stars\cite{motz},
this is an upper limit to the true angular momentum.},
 $I=\frac{2}{5} M_{core} R_{core}^2$.
Using 
\ba
\Omega_K &=& \sqrt{\frac{GM_o}{R_o^3}}
\ea
we get 
\ba
J_{core}= \frac{2}{5} M_{core} R_{core}^2 \sqrt{\frac{GM_o}{R_o^3}}\label{jcore}
\ea
As in the previous merger case, a part of the core angular 
momentum($x$) 
and 
the core mass($y$) collapses into the black hole($M_{BH} = yM_{core}$).  
Then we get 
\ba   
J_{BH} &=&xy J_{core}\\
&=& xy \frac{2}{5} M_{core} R_{core}^2 \sqrt{\frac{GM_o}{R_o^3}}\\
&=& \tilde{a}(yM_{core})^2\frac{G}{c}.
\ea
Then  the black hole angular momentum parameter is given by
\ba
\tilde{a} &=&  \frac{2}{5} \frac{x}{y} \frac{R_{core}^2c}{GM_{core}}
 \sqrt{\frac{GM_o}{R_o^3}}\\
&=& \frac{2}{5} \frac{x}{y} (\frac{R_{core}}{R_o})^{3/2} 
\sqrt{\frac{R_{core}c^2}{GM_{core}}}
 \sqrt{\frac{M_o}{M_{core}}}
\ea
With 
\ba
R_o \sim 10^5 km, \ \ , R_{core} \sim 10^3 km \nonumber\\
M_o \sim 15 M_{\odot}, \ \ ,  M_{core} \sim 2 M_{\odot} (\sim 3 km)\label{core}
\ea
we get
\ba
\tilde{a} \sim  0.1 \frac{x}{y} 
\ea
This depends strongly on the 
numbers taken in eq.(\ref{core}) and on how the core angular momentum
is determined. The specific angular momentum of the core can be calculated 
using  
eq.(\ref{jcore}) and eq.(\ref{core}) as
\ba
a_{core} &=& J_{core}/M_{core} = \frac{2}{5} R_{core}^2 
\sqrt{\frac{GM_o}{R_o^3}}\\
         &=& 1.8 \times 10^{14} cm^2 s^{-1}
\ea
which is much smaller than that from the numerical simulation of 
collapsars\cite{popham}: $a_{core} \sim 10^{16} cm^2 s^{-1}$.  
Of course, if the core angular momentum is not completely redistributed
during the precollapse evolution the core is likely to be spinning faster 
than $\Omega_K$; The maximum possible value  comes from replacing $\Omega_{K}$
determined at the progenitor radius in  eq.(\ref{jcore}) 
 by the 
one determined at the core radius
\ba
\Omega_{K}^{core} &=& \sqrt{\frac{GM_{core}}{R_{core}^3}}\\
           &=& (1.3 \times 10^2) \Omega_K,
\ea
and we get
\ba
a_{core} = 2.34\times 10^{16} cm^2 s^{-1}.
\ea
Then the angular momentum parameter can be given by   
\ba
\tilde{a} = \frac{x}{y} \frac{a_{core}}{a_{max}} = 2.3\frac{x}{y}
\ea
  
In summary, it is very plausible to  have a rapidly rotating black 
hole($ \tilde{a} > 0.1$)
 as a 
resulting  object in the center in the merging 
systems and also in hypernovae of large angular momentum progenitors, but
a precise value of $\tilde{a}$ will be difficult to calculate. 

\section{Magnetized accretion disks }

A black hole by itself cannot keep   magnetic fields on it for a long time.   
Magnetic fields  diffuse away 
in a short time  $\sim R_H/c$\cite{TPM}\cite{lop}. The most plausible 
environments which  
can support a magnetic field  threading  
the black hole  are 
accretion disks
surrounding the black hole. There are two issues about accretion disks 
that we need to consider in order to decide whether the 
Blandford-Znajek process is a viable power source for
gamma ray bursts: The life 
time should be long enough to extract the bulk of the black hole spin energy 
and also the magnetic field on the disk should be 
strong enough  to power the gamma ray bursts from the spinning 
black hole.
A strong magnetic field  on the inner part of 
the accretion disk is neccessary to keep the magnetic pressure comparable to 
that of magnetic fields on the black hole\cite{TPM}.  It has been also 
demonstrated\cite{ghosh} from the axisymmetric solutions discussed 
in section 4. Since the magnetic field on the disk affects the angular 
momentum transfer of the accretion disk via magnetic braking\cite{blandford}
and/or magnetic viscosity, the accretion process  
is not independent of the magnetic field on the disk.
\noindent  The magnetic field on the disk should not be  larger than the value
from the  equipartition argument:
\ba
\frac{B_{eqp}^2}{8\pi} \sim P_{disk};\label{eqipb},
\ea
and $B_{eqp}$ can be considered as an upper limit to the magnetic field 
which can be 
supported by the accretion disk.
Recent numerical calculations on the hyper-accreting black hole by Popham et 
al.\cite{popham} show 
\ba
P_{disk} \sim 10^{30} erg/cm^3\\
B_{eqp} \sim 5\times10^{15} gauss,
\ea      
which implies the accretion disk of the hyper-accreting black hole may support 
a magnetic field strong enough for the gamma ray burst.  However it  depends on 
the detailed  
mechanism how  strong  a magnetic field can be built up on the disk.  One of 
the approaches is that the magnetic field is evolved  magnetohydrodynamically
 during the accreting process, which   depends on the magnetic 
viscosity. 

The magnetic viscosity, 
$\nu^{mag}$, is defined\cite{magvis} as
as   
\ba
\frac{B_rB_{\phi}}{4\pi} = - \nu^{mag} \, (r\frac{d\Omega_{disk}}{dr}) \, 
\rho, \label{vismag}     
\ea
which can be parametrized by the viscosity parameter, $\alpha$,   in 
terms of dimensionful quantities of the disk as\cite{SS}
\cite{pringlea},
\ba
\nu^{mag} = \alpha^{mag} \, c_s \, H \label{alphav}
\ea
where $c_s$ is the sound velocity of the disk($c_s = 
\sqrt{\gamma P_{disk} / \rho}$) and $H$ is a half of the disk thickness.
Using hydrostatic equilibrium perpendicular to the disk plane,
\ba
H &=& \sqrt{\frac{P_{disk}}{\rho}}/\Omega_{disk}\\
  &=& \frac{c_s}{\Omega_{disk} \sqrt{\gamma}}\label{homega}
\ea
we get
\ba
\nu^{mag} = \alpha^{mag} \, \frac{c_s^2}{\Omega_{disk} \sqrt{\gamma}}.
\ea
For a Keplerian orbit 
\ba
r\frac{d\Omega_{disk}}{dr} = - \frac{3}{2} \Omega_{disk},\label{domega}
\ea
eq.(\ref{vismag}) can be written as
\ba
\frac{B^{dyn}_rB^{dyn}_{\phi}}{4\pi}= \frac{3}{2} \alpha^{mag} \, P_{disk} 
\sqrt{\gamma}
\label{vismagal}  
\ea
We can see that the Maxwell stress of the magnetic field which has been built 
up by the accretion 
dynamo is also proportional to the disk pressure but for a 
different reason from 
that of the equipartition argument, eq.(\ref{eqipb}). Numerical estimations
of the viscosity parameter, $\alpha^{mag}$, obtained for  various boundary 
conditions range from 0.001 to 0.005\cite{magvis}. This  means that the 
magnetic pressure is only a small fraction of the disk pressure\cite{lop}. 
Using the estimation of 
disk pressure by Popham et al.\label{popham}, we can estimate the dynamically 
generated magnetic field by the accretion: 
\ba
B^{dyn} \sim 10^{13} gauss
\ea
which may not be strong enough for a  gamma ray bursts powered by 
the Blandford-Znajek process.

However, an accretion dynamo might  not be the only process responsible for 
the 
magnetic fields on 
the disk. We know from the recent observations that there are a number of 
pulsars,
magnetars, which are believed to have  a strong magnetic field of 
$\sim 10^{15} $ G.  Although the origin\cite{kr} of 
such strong 
magnetic fields
 is not  
well known at the moment,   one may  consider the 
case that a debris torus or disk around the black hole which was formed by 
the disruption of a neutron star retain the high ordered field of 
that neutron star.

Now consider the axisymmetric solution around a disk with a force-free 
magnetosphere which has been discussed by Blandford\cite{blandford}. Here we
adopt  cylindrical coordinates, $(r,\phi,z)$, where the $z-$direction is 
perpendicular to the disk.
The sum of current flows into the disk up to radius $r$ defines 
the surface current density,   
$J_r$, which is proportional to the poloidal component of the magnetic field on
the disk, $B_{\phi}$:
\ba
\frac{4\pi J_r}{2} = -  B_{\phi}\label{jr}
\ea
where 2 in the denominator comes from the fact that the radial current density
$J_r$ includes  the 
currents into the disk both from  above and below. 
The torque exerted by the  annular 
ring with width $\Delta r$ of the disk due to 
the Lorentz force is given by
\ba
\Delta T &=& - r \, 2\pi r J_r \, B_z \Delta r\\
         &=& 2r\frac{B_{\phi}B_z}{4\pi}\Delta S, 
\Delta S = 2\pi r \Delta r\label{deltatd} 
\ea
For the steady state accretion disk with surface density, $\Sigma$, 
the angular momentum conservation
can be written by
\ba
\Sigma v_r \frac{\partial(r^2 \Omega)}{\partial r} 2\pi r \Delta r &=&
\Delta T + \frac{\partial G}{\partial r} \Delta r\label{anguld}\\
\dot{M} \frac{\partial(r^2 \Omega)}{\partial r}  \Delta r &=&
\Delta T + \frac{\partial G}{\partial r} \Delta r\label{anguld1}
\ea
where the torque due to the shear force of differential rotation, 
$G$\cite{pringlea}, is given by
\ba
G = 2 \pi r \,  \nu \Sigma (r \frac{\partial \Omega}{\partial r}) \, r\label{G}
\ea
To see the effect of magnetic field,   we consider only 
the magnetic viscosity for the moment.  Using 
eq.(\ref{vismag}), we get
\ba
G^{mag} &=&  2\pi r^2 \frac{\Sigma}{\rho}\frac{B_{\phi}B_r}{4\pi}\\
        &=&  4\pi r^2 \, H \, \frac{B_{\phi}B_r}{4\pi}\\
        &=& 4\pi r^2 \frac{c_s}{\Omega_{disk}}\frac{B_{\phi}B_r}{4\pi}
\ea
where $2H \, \rho = \Sigma$.
  Assuming a simple power dependence of 
$H\frac{B_{\phi}B_r}{4\pi} \propto  r^n$, 
\ba
\frac{\partial G^{mag}}{\partial r} = 4(2+n)\pi r H \frac{B_{\phi}B_r}{4\pi}
\ea
and we get from eq.(\ref{anguld1})
\ba
\dot{M} \frac{\partial (r^2 \Omega_{disk})}{\partial r} = 
r^2B_{\phi}B_z[1+ (2+n)\frac{H}{r}\frac{B_r}{B_z}]
\ea      
Since $H \ll r$, the accretion rate is determined by the magnetic braking 
as far as $B_r$ is not much larger than $B_z$ and the second term in the 
right hand side of the above equation can be 
neglected.  For $r \gg GM_{BH}/c^2$
where the disk angular velocity can be approximated by a Keplerian velocity,
 $\partial(r^2\Omega)/ \partial r = r\Omega_{disk}/2$,
\ba
\dot{M} = 2rB_{\phi}B_z/\Omega_{disk}\label{mdot}
\ea 
Using the axisymmetric solution\cite{blandford},
\ba
B_{\phi} = 2r\Omega_{disk} B_z/c\label{bbz}
\ea
we get 
\ba
\dot{M} = 4 \, r^2 B_z^2/c\label{mdot1}
\ea
Since for a steady accretion $\dot{M}$ is independent of $r$, $B_z \sim 1/r$ 
at large distance. For a numerical estimation, we take $ r = 10^6 cm$ and 
$B_z = 10^{14} gauss$  Then we 
get
\ba
\dot{M} = 6 \times 10^{-4} \,  
(\frac{B_z}{10^{14}gauss})^2 \,  M_{\odot} \,s^{-1}
\ea
which is  much larger than Eddington luminosity, $\dot{M}_{Ed} 
\sim 10^{-16} \, \frac{M_{BH}}{M_{\odot}}\, M_{\odot}s^{-1}
$.  The total accretion 
during the gamma ray burst period is at most $10^{-1} M_{\odot}$, 
which may not
change the discussions on Blandford-Znajek process in the previous sections.
Hence black holes 
surrounded by the 
hyper-accretion 
disks  might be a good candidate for the central engine of the gamma ray bursts.  
The magnetic field on the disk  extracts  not only the angular momentum as  described 
above but also a substantial energy out of disk as it does on the black 
hole. The power of the disk magnetic braking can be calculated using 
eq.(\ref{deltatd})
\ba
\Delta P_{disk} &=& \Omega_{disk} \Delta T\\
    &=& 2 \, \frac{B_{\phi}B_z}{4\pi} (r\Omega_{disk}) \Delta S
\ea
Using eq.(\ref{bbz}) with Keplerian angular velocity of the disk , we get
\ba
\Delta P_{disk} &=& 4 \, c \, \frac{B_z^2}{4\pi} 
(\frac{GM_{BH}}{rc^2}) \Delta S\\
dP_{disk} & =& \frac{1}{\pi c} B_z^2r^2 \frac{GM_{BH}}{r^3} dS,\label{PDISK}
\ea
then setting $dS = 2\pi rdr$ and  using the steady-state relation
$B_z^2 r^2 = B_z(r_{in})^2 r_{in}^2$, we can obtain the total power by 
integrating eq.(\ref{PDISK})  from
$r_{in}$ to infinity, 
\ba
 P_{disk} = 2B_z(r_{in})^2 r_{in} \frac{GM_{BH}}{c}\label{pdiskf}
\ea
where $r_{in}$ is the distance from the black hole to the inner edge of the
accretion disk. 
Compared to the BZ power from the  spinning black hole, 
 eq.(\ref{power0})),
the ratio can be given by
\ba
\frac{P_{disk}}{P_{BH}} 
= 8 \frac{r_{in} c^2}{GM f(h) \tilde{a}^2}  
(\frac{B_z(r_{in})}{B_H})^2
\label{dBH1}
\ea

The current conservation condition, 
namely that the total current flows onto the black hole 
should go into the inner edge($r_{in}$) of the accretion disk, implies
\begin{equation}
2M B_{\phi}^H(\theta=\pi/2) = \tilde{\omega}(r_{in})B^{disk}_{\phi}(r_{in}).
\end{equation}
 Since the cylindrical radius in Kerr geometry $\tilde{\omega}(r_{in})
> 2M $, we can see that $B_{\phi}^H$ is larger than
$ B^{disk}_{\phi}$.
 From the boundary conditions on the horizon\cite{TPM} in the optimal case,
\begin{equation}
B_{\phi}^H = -\Omega_H M B_H,
\end{equation}
and on the accretion disk\cite{blandford} with angular
velocity $\Omega_D$, 
\begin{equation}
 B_{\phi}^{disk} =
-2 \Omega_D r B_z^{disk},
\end{equation}
we get
\begin{equation}
\frac{B_z(r_{in})}{B_H} =  \sqrt{\frac{GM}{r_{in}c^2}}
\frac{\tilde{a}}{2}\frac{G M}{r_Hc^2} < 1,
\end{equation}
and the power of the disk  magnetic braking $P_{disk}
= \frac{2}{f(\tilde{a})}(\frac{GM}{r_Hc^2})^2 P_{BZ}$.
It shows that the magnetic field on the horizon cannot be smaller than
that on the inner edge of the accretion disk\cite{li} and 
the disk power need not
be substantially larger than that from the black hole\cite{lop}
\cite{ghosh}.

\section{Conclusion}

We have evaluated the power and energy that can be extracted from
a rotating black hole immersed in a magnetic field, the 
Blandford-Znajek effect. We improve on earlier calculations to find that
the power from a black hole of given mass immersed a an external field is
ten times greater than previously thought. The amount of energy that can
be extracted from a black hole in this way is limited by the fact that
only 29\% of the rest mass of a black hole can be in rotational energy,
and that the maximum efficiency with which energy can be extracted from
the hole via the Blanford-Znajek effect is 31\%. The net amount of energy
that can be extracted is therefore 9\% of the rest energy of the black
hole. We consider various scenarios for the formation of rotating black
holes in gamma-ray burst engines, and while the resulting angular momenta
are quite uncertain in some cases, it seems that the required values of
the rotation parameter, $\tilde{a} \gsim 0.5$, are achievable.

The rate at which angular momentum is extracted depends on the magnetic
field applied to the hole. A field of $10^{15}$\,G will extract the
energy in less than 1000\,s, so time scales typical of gamma-ray bursts
can be obtained.  Since the black hole cannot carry a field, there must
be an ambient gas in which the field is anchored that drives the Poynting
flux from the black hole. The most obvious place for it is the accretion
disk or debris torus surrounding the black hole just after it formed.
                  
It has been argued  that a field turbulently generated
in the disk would not give a strong Blandford-Znajek flux\cite{lop}, because
the disk would dominate the total Poynting output, and no more than the
disk's binding energy could be extracted. We show explicitly that a field
in the disk could be much greater, for example if it is derived from the
large, ordered field of a neutron star that was disrupted. The
field distribution proposed by Blandford\cite{blandford} for such a case 
would allow
the field on the hole to be much greater than on the disk, such that the
Poynting flow would not be dominated by the disk  and not subject to any
obvious limits imposed by the disk.

We also note that a Poynting flow may provide an alternative way of providing
a very large magnetic field for the shocked material that radiates the
afterglow and the gamma-ray burst itself: the standard assumption is that
the required high fields grow turbulently in the shocked gas, up to
near-equipartition values. But the field in the Poynting flow only decreases
as $1/r$, so if it is $10^{15}$\,G at $r=10^5$\,cm, it could be as high as
$10^{4}$\,G at the deceleration radius ($10^{16}$\,cm), ample to cause an
energetic gamma-ray burst.

\noindent {\bf Acknowledgement}

We would like to thank Roger Blandford, Sterl Phinney and Kip Thorne
for guidance and useful discussions. HKL is very grateful to the Nuclear Theory 
Group, SUNY at Stony Brook for the hospitality during his one year stay, where
the most of collaborated works on this paper has been done..  
This work is supported by the U.S. Department of Energy 
under Grant No. DE-FG02-88ER40388.
HKL is also supported in part by KOSEF-985-0200-001-2 and BSRI-98-2441. 

\newpage

\noindent {\bf Appendix 1. Rotating Black Hole}

A rotating black hole is defined as its angular momentum($J$) and mass($M$). 
The angular momentum is measured by the non-newtonian gravitational effect
(gravitomagnetic effect) far from the black hole and the mass is measured 
by the newtonian gravitational field far from the black hole. The solution 
of Einstein's equation\cite{kerr} defines the metric around the rotating 
black hole with specific angular momentum $a=J/M$. Using the
 Boyer-Lindquist coordinates\cite{bl}, in the    
 natural unit $G=c=1$ the metric can
be written as
\ba
ds^2 = -\alpha^2 dt^2 + g_{ij}(dx^i + \beta^i dt)(dx^j + \beta^j dt).\label{ds}
\ea
$\alpha$ and $\beta$ are lapse function and shift function\cite{TPM}
defined respectively 
as
\ba
\alpha &=& \frac{\rho \sqrt{\Delta}}{\Sigma} \label{lapse}\\
\beta^{\phi} &=& - \frac{2aMr}{\Sigma^2} \label{beta}\\
\beta^{\theta} &=& \beta^{r} = 0\\
\ea
where
\ba
\Delta &=& r^2 + a^2 -2Mr\label{delta}\\
\rho^2 &=& r^2 + a^2 \cos^2 \theta\label{rho}\\
\Sigma^2 &=& (r^2 + a^2)^2 - a^2 \Delta \sin^2\theta \label{sigma}\\
\ea
The metric tensor $g_{ij}$ is given by
\ba
g_{rr} &=& \frac{\rho^2}{\Delta}, \,\, g_{\theta\theta} = \rho^2,
\,\,\, g_{\phi\phi} =
\tilde{\omega}^2\label{gij}
\ea
where
\ba
\tilde{\omega} &=& \frac{\Sigma}{\rho} \sin \theta\label{tomega}
\ea
The volume and area elements $dV$ and $d\vec{S}$ are defined in the 
standard way\cite{TPM} by
\ba
dV &=& \sqrt{det(g_{ij})} dr d\theta d\phi\\
   &=& \frac{\rho^2 \tilde{\omega}}{\sqrt{\Delta}} dr d\theta 
       d\phi\label{dV} \\
dS_i &=& \sqrt{det(g_{ij})} \epsilon_{ijk} \frac{\partial x^j}{\partial b}
 \frac{\partial x^k}{\partial c} db \, dc \label{dsi}
\ea
The circumference of a circle around the axis of symmetry is $2\pi 
\sqrt{g_{\phi\phi}} = 2\pi \tilde{\omega}$.
 
\noindent The horizon, $r_H$, is defined as a larger root of 
$\Delta_{r=r_{H}} = 0 $,  
\ba
r_H &=& M + (M^2 - a^2)^{1/2}\label{rh}\\
2Mr_H &=& r_H^2 + a^2 \label{rh1}
\ea
 
\noindent The surface area of the horizon, $A_H$,  is given by
\ba
A_H = 4\pi(r_H^2 + a^2)\label{area}
\ea
and the entropy and the irreducible mass of the black hole are given 
respectively by
\ba
S_H &=& \frac{k_B}{4  \hbar} A_H\\
    &=& \frac{\pi k_B}{\hbar}(r_H^2 + a^2)=  \frac{2\pi k_B}{\hbar}Mr_H.
\label{entropy}\\
M_{irr} &=& \sqrt{\frac{A_H}{16\pi}} = \frac{1}{2}\sqrt{r_H^2 + a^2}= 
\sqrt{\frac{S_H}{4\pi}}\label{mirr}
\ea
The the mass of the black hole can be rewritten in terms of $J, M_{irr}$, 
and $S_H$,
\ba
M &=& \sqrt{\frac{S_H}{4\pi} +\frac{J^2}{S_H}} = 
\sqrt{M_{irr}^2 + \frac{J^2}{4M^2_{irr}}}\label{mmirr}
\ea
The angular velocity of the black hole is given by
\ba
\Omega_H &=& -\beta^{\phi}_H = \frac{a}{2Mr_H} = \frac{J}{2M^2r_H}\\
&=& \frac{J}{2M^3(1+\sqrt{1-J^2/M^4})} = \frac{1}{2M} \frac{\tilde{a}}
{1+\sqrt{1-\tilde{a}^2}}
\ea   
where $\tilde{a}$ is the  angular momentum parameter defined as 
$\tilde{a} = J/M^2$.

\newpage

{\bf Appendix 2. Electromagnetic fields in vaccum arround rotating black hole}
 
For a rotating black hole immersed in an asymptotically uniform magnetic field
\ba
\vec{B} = B \hat{z}, \,\,\, r \rightarrow \infty, \label{binfty}
\ea
the poloidal components of electric and magnetic fields can be written
analytically \cite{TPM}
\ba
B_{r} &=& \frac{B}{2\Sigma \sin \theta} \frac{\partial X}{\partial \theta}
\label{ebanal1}\\
B_{\theta} &=& -\frac{B\sqrt{\Delta}}{2\Sigma \sin \theta}
\frac{\partial X}{\partial r}\\
E_r &=& \frac{-Ba\Sigma}{\rho^2}[\frac{\partial \alpha^2}{\partial r} +
\frac{M\sin^2\theta}{\rho^2}(\Sigma^2 -4a^2Mr)\frac{\partial}{\partial r}
(\frac{r}{\Sigma^2})\\
E_{\theta} &=& \frac{-Ba\Sigma}{\rho^2\sqrt{\Delta}}
[\frac{\partial \alpha^2}
{\partial \theta} +
\frac{\sin^2\theta}{\rho^2}(\Sigma^2 -4a^2Mr)\frac{\partial}{\partial \theta}
(\frac{1}{\Sigma
^2})]. \label{ebanal}
\ea   
where $X= \frac{\sin^2\theta}{\rho^2}(\Sigma^2 -4a^2Mr)$ and $a$ is
the specific
angular momentum of the black hole, $J = aM$. Here we adopt the natural units
$G=c=1$.

The radial components of magnetic field
on the horizon, $B^H_r$,   can be written  explicitly by
\ba
B^H_r &=& \frac{B\cos\theta}{(r_H^2 +(a/c)^2 \cos^2 \theta)^2}\\ \nonumber
& & \ \ \ \ [(r_H^2 -(a/c)^2)
(r_H^2 -(a/c)^2 \cos^2 \theta) + 2(a/c)^2 r_H (r_H-M)
(1 + \cos^2 \theta)]\label{bhr}\\
E^H_r &=& - \frac{aB}{2r_Hc(r_H^2 +(a/c)^2 \cos^2 \theta)^2}\\ \nonumber
 & & \ \ \ \ (r_H^2 -(a/c)^2)
[(3r_H^2 + (a/c)^2(1+\cos^2\theta))
\cos^2\theta -r_H^2]\label{ehr}\\
\ea    
 It is clear that $B^H_r$ vanishes as the rotation of black hole approaches
extreme
rotation
\ba
a \rightarrow GM/c, \,\,\,
r_H \rightarrow GM/c^2
\ea
This is what has been observed \cite{king}\cite{bicak} as the absence of
the magnetic
flux across the maximally rotating black hole.
There are no $\phi$  components of electric and magnetic fields in
 eq.(\ref{ebanal}),
and the $\theta$ components
vanish as $\sqrt{\Delta}$ as the horizon is approached.
The main reason for these vanishing tangential components is
that there are no currents outside the horizon in  charge-free space.
In the absence of in/out currents there is no
 toroidal magnetic field on
the horizon.
This also requires a vanishing tangential component of electric field on the
horizon,  which induces
a charge separation on the horizon\cite{TPM}.
Therefore there is  no current on which the external magnetic field
exerts torques
to slow the black hole
down  so as to extract its rotational energy.
One can verify from eq.(\ref{ebanal1}) - eq.(\ref{ebanal}) that there is
no outward component of the Poynting vector
(essentially it is due to the absence
of the $\phi$ component of the magnetic field), which means no outflow of
energy to infinity.

\newpage
 
\noindent {\bf Appendix 3. Axial-symmetric Force-free Magnetosphere}

In this appendix, the original formulation\cite{BZ} of the  
Blandford-Znajek process
is summerized. 
Consider the stationary and axial symmetric situation where all the
partial derivatives of the physical quantities with respect to time and 
azimuthal angle $\phi$ are
vanishing.
The electromagnetic field tensor is given by the
vector potential $A_{\mu}$ as
\ba
F_{\mu\nu} &=& A_{\nu,\mu} -  A_{\mu,\nu}\label{fmunu}\\
\ea
and the magnetic field is given by
\ba
B_r &=& F_{\theta\phi} = A_{\phi,\theta} \\
B_{\theta} &=& F_{\phi r} = -A_{\phi, r}\\
B_{\phi} &=&  F_{r \theta}= A_{\theta,r} - A_{r,\theta}
\ea   
The magnetic flux, $\Psi$, through a circuit encircling 
 $ \phi = 0 \rightarrow 2 \pi$,
\ba
\Psi
= \oint  A_{\phi} d\phi = 2\pi A_{\phi} \label{psiaphi}
\ea
It defines a magnetic surface on 
which  $A_{\phi}(r,\theta) $ is constant and therefore is characterized by the 
magnetic flux $\Psi$ contained inside it.   Magnetic field lines are 
spiraling on 
the surface and no magnetic field lines can cross the magnetic surface. 

\noindent {\bf (1) Force-free magnetosphere}

The force-free condition for the  magnetosphere with the current density
$J^{\mu}$,
\ba
F_{\mu\nu}J^{\mu} = 0, \label{forfree}
\ea    
can be written by
\ba
A_{0,r}J^r + A_{0,\theta}J^{\theta} &=& 0 \label{1}\\
A_{\phi,r}J^r + A_{\phi,\theta}J^{\theta} &=& 0 \label{2}\\
-A_{0,r}J^0 - A_{\phi,r}J^{\phi} &=& B_{\phi}J^{\theta}\label{3}\\
A_{0,\theta}J^0 + A_{\phi,\theta} J^{\phi} &=& B_{\phi}J^r. \label{4}
\ea 
>From eq.(\ref{1}) and (\ref{2}) we get
\ba
\frac{A_{0,r}}{A_{0,\theta}} = \frac{A_{\phi,r}}{A_{\phi,\theta}}
 = -\frac{J^{\theta}}{J^r}
\ea
We can see that $A_0 $ is also constant along the magnetic field
 lines
and the electric field is always perpendicular to the magnetic 
surface.
We can define a function $\omega(r,\theta)$, 
\ba
\frac{A_{0,r}}{A_{\phi,r}}&=& \frac{A_{0,\theta}}{A_{\phi,\theta}}
=- \omega \, \label{omega}\\
dA_0 &=& -\omega dA_{\phi},
\ea
which is also constant along the magnetic surface. $\omega$ can be 
identified as an electromagnetic angular velocity(an angular velocity
 of magnetic field line on a 
magnetic surface).
We can also suppose a function $\mu(r,\theta)$ satisfying
\ba
4 \pi J^r &=& \frac{\mu}{\sqrt{g}} 
A_{\phi, \theta}\nonumber \\  
4 \pi J^{\theta}&=& \frac{\mu}{\sqrt{g}}
A_{\phi,r }\label{jtheta}
\ea
where $\sqrt{g} = \sqrt{-det(g_{\mu\nu})} = \rho^2 \sin \theta$.  The current 
conservation
\ba
\frac{1}{\sqrt{g}} ( \sqrt{g}
 J^{\mu})_{,\mu}
=0\label{Jconserv}
\ea
leads to 
\ba
(\mu A_{\phi,\theta}),r = (\mu A_{\phi,r})_{,\theta}
\ea
which implies that $\mu$ is also constant on the magnetic surface.

From eq.(\ref{3})(or equivalently from eq.(\ref{4})), we get
\ba
J^{\phi} = \omega J^0 + \frac{\mu}
{4 \pi \sqrt{g}}
 B_{\phi}
\ea
Together with eq.(\ref{jtheta}), we now have 
an expression of the current 
in terms of the magnetic field and charge density with yet 
undetermined function $\omega$ and $\mu$.   
\noindent The  outward current $I$ can be calculated from the $r$ and 
$\theta$ components of the current.  Since the outward curent is
proportinal to the magnetic field, the current between the magnetic 
surface,$dI$, can be written in terms of the magnetic flux as
\ba
 dI &=& \frac{1}{4\pi}  \mu  d\Psi\\
 &=&  \frac{1}{2} \mu d A_{\phi}\label{deltaI}
\ea

\noindent {\bf (2) Energy and Angular Momentum Outflow}

For a stationry and axial symmetric system, the conserved fluxes can 
be defined about the axis of symmetry.   The force-free condition 
ensures the conserved  electromagnetic energy flux 
\ba
{\cal E}^{\mu} = T^{\mu}_0
\ea
and angular momentum flux
\ba 
{\cal L}^{\mu} = -T_{\phi}^{\mu}
\ea
where the electromagnetic energy momentum tensor is given by
\ba
T^{\mu\nu} = \frac{1}{4\pi}(F^{\mu}_{\, \rho}F^{\nu \rho} - \frac{1}{4} 
g^{\mu\nu}
F_{\rho\sigma}F^{\rho\sigma}).\label{tmunu}
\ea
The  energy flux and angular momentum flux 
have a simple relation,
\ba
\cal{E}^{\mu} = \omega \cal{L}^{\mu}
\ea
and the poloidal components can be written as
\ba
{\cal E}^{r} &=& -\frac{1}{4\pi}\omega  \frac{\Delta}{\rho^4}
A_{\phi,\theta}B_{\phi}\label{er}\\
{\cal E}^{\theta} &=& \frac{1}{4\pi} \omega  \frac{\Delta}{\rho^4}
A_{\phi,r}B_{\phi}\label{etheta}
\ea
Using the physical component, ${\cal E}_{\hat{r}}$,
\ba
{\cal E}_{\hat{r}} 
= \sqrt{g_{00}/g_{rr}}{\cal E}_r =  \sqrt{g_{00}g_{rr}}{\cal E}^r
\ea
we get the power at infinity given by
\ba
P &=& \int {\cal E}_{\hat{r}} \cdot dS_{\hat{r}}\\
  &=& -\frac{1}{4\pi} \int \omega \sqrt{\frac{g_{00}}
      {g^2_{\theta \theta}g_{rr}}} B_{\phi}
      A_{\phi,\theta} \sqrt{g_{\theta \theta}g_{\phi \phi}} 
      d\theta d \phi \\
 &=& -\frac{1}{4\pi} \int \omega \sqrt{\frac{g_{00}g_{\phi \phi}}
     {g_{rr}g_{\theta \theta}}} B_{\phi}
      A_{\phi,\theta} d\theta d \phi \\
  &=& - \frac{1}{4\pi} \int \omega 
      (\frac{\Delta \sin \theta}{\rho^2}B_{\phi})
      A_{\phi,\theta} d\theta d\phi \label{P1}
\ea
Defining $B_T$\cite{BZ} as
\ba
B_T &=& \frac{\Delta \sin \theta}{\rho^2}B_{\phi}\label{BT}
\ea
the power can be written as
\ba
P &=& - \frac{1}{2}  \int \omega B_T A_{\phi,\theta} 
      d\theta \label{P11}\\
  &=& - \frac{1}{2}  \int \omega B_T dA_{\phi}.\label{P2}
\ea

From the inhomogeneous Maxwell equations
\ba
F^{\mu\nu}_{\, \, ; \nu} = 4 \pi J^{\nu}\label{fmununu}
\ea
we get 
\ba
B_{T,\theta} &=& 4 \pi\sqrt{g}J^r\\
B_{T,r} &=& -  4 \pi \sqrt{g}J^{\theta}\label{BTRtheta}
\ea
Using eq.(\ref{jtheta}),   
\ba
\mu A_{\phi,\theta} &=& B_{T,\theta}\nonumber\\
\mu A_{\phi,r} &=& B_{T,r}
\ea
which shows that $B_T$ is also constant along the magnetic surface. Since
\ba
\mu dA_{\phi} = dB_T,
\ea
the outward 
current can be calculated as
\ba
dI &=& \frac{1}{2}  \mu dA_{\phi}\label{dida}\\
   &=& \frac{1}{2}  dB_T\label{didbt}
\ea
and we get 
\ba
 I(\theta)  = 
\frac{1}{2} \int  dB_T 
    = \frac{1}{2} B_T(\theta)\label{Ibt}
\ea
where $B_T(\theta=0) =0$ has been used. Using the physical component
, $B_{\hat{\phi}} = \frac{B_{\phi}}{\sqrt{g_{rr}g_{\theta \theta}}}$,
eq.(\ref{Ibt})  can be written
\ba
I = \frac{1}{2} \tilde{\omega} \alpha B_{\hat{\phi}}
\ea
which is nothing but the Ampere's law on the stretched horizon
\ba
2\pi \tilde{\omega} B^{H}_{\hat{\phi}} = 4 \pi I
\ea
using the tangential field on the stretched horizon\cite{TPM}
\ba
 B^{H}_{\hat{\phi}} =  \alpha B_{\hat{\phi}}
\ea   

\noindent Using eq. (\ref{P2}), (\ref{dida}) and (\ref{Ibt})
we get
\ba
dP &=&  -  \omega I dA_{\phi}\\
 &=& - \frac{1}{2\pi} \omega I d\Psi\\
P  &=& - \frac{1}{2\pi} \int \omega I d\Psi \label{ppsi}
\ea
Since $\omega, I$ and $\Psi$ are constant along the magnetic 
surface we can evaluate the integral on the horizon of the 
black hole:
\ba
P  &=& - \frac{1}{2\pi} \int_{horizon} \omega I d\Psi \label{ppsih}
\ea
Hence  we can see  that it is a proof of  the 
simple-minded
derivation of eq.(21) in the text with $\omega$ 
identified with $\Omega_F$.        

\newpage 

\noindent {\bf Appendix 4. Power from Rotating Black Hole}

Using the boundary condition\cite{z77} that $B_r$ is finite
and 
\ba
B_T &=& \frac{\sin \theta [\omega(r_H^2 + a^2) -a]}{r_H^2 + 
        a^2 \cos\theta}B_r\label{btbrz}\\
    &=& \frac{\sin \theta(\omega-\Omega_H)\Sigma_H}{\rho_H^2}
        B_r\label{btbrl}
\ea
which can be written for $B_{\phi}$ as
\ba
B_{\phi} &=& \frac{\rho^2}{\Delta}\frac{(\omega-\Omega_H)\Sigma_H}
             {\rho^2}B_r\\
         &=& \frac{(\omega-\Omega_H)\Sigma_H}{\Delta} B_r\\
         &=&  \frac{(\omega-\Omega_H)\Sigma_H}{\Delta}
              \rho\tilde{\omega} B_{\hat{r}}\label{bphihatr}
\ea
and the toroidal component on the stretched horizon is
\ba
 B^{H}_{\hat{\phi}} = (\omega -\Omega_H)\tilde{\omega} B_H
\ea
where $B_H = B_{\hat{r}}$.
Then the current  up to $\theta = \theta_H$
can be written using eq.(\ref{Ibt})
\ba
I(\theta_H) = \frac{1}{2} \frac{\Delta}{\rho^2} \sin \theta B_{\phi}
      =  \frac{1}{2} (\omega-\Omega_H) \tilde{\omega}^2 
           B_H\label{Ibr}
\ea
which is exactly the same current derived in the membrane 
paradigm\cite{TPM}.   
 
The total power out of the black hole can be calculated using eq.
(\ref{P11}), (\ref{Ibt}), and (\ref{Ibr}),
\ba
P &=& \frac{1}{2}  \int \omega B_T B_r d\theta\\
  &=& - \int \omega I \rho \tilde{\omega} B_H  d\theta \\
  &=& - \frac{1}{2} \int \omega (\omega - \Omega_H) B_H^2
      \rho \tilde{\omega}^3 d\theta  \label{pbtbr}
\ea
Adopting the
optimal condition\cite{TPM}, $\omega \sim \Omega_H/2$, the power can be
written as
\ba
P &=& -\frac{1}{2} \int \omega (\omega - \Omega_H) B_H^2
      \rho \tilde{\omega}^3 d\theta\\ 
 &=&   \frac{1}{8}(\frac{a}{2Mr_H})^2 
     \int \frac{(r_H^2 + a^2)^3 \sin^3 \theta}
   {r_H^2 + a^2 \cos^2 \theta} B_H^2 d\theta \\
 &=&    \frac{1}{32}(\frac{a}{M})^2 \frac{(2Mr_H)^2}{r_H^2}
   (1+(\frac{r_H}{a})^2) 
    \int \frac{\sin^3 \theta}
   {(\frac{r_H}{a})^2 + \cos^2 \theta} B_H^2 d\theta
\ea
Using the integral identity,
 \ba
\int_0^{\pi/2} \frac
    {\sin^3\theta}
   {(\frac{r_H}{a})^2 + \cos^2\theta} d\theta = [(h+1/h) \arctan h -1]
\ea
where $ h$ is defined in ref \cite{ok} as
\ba
h &=& \frac{a}{r_H} \\
  &=& \frac{J/M^2}{1+ \sqrt{1-J^2/M^4}} = \frac{\tilde{a}}{1+
\sqrt{1-\tilde{a}^2}},\\  
\frac{M}{r_H} &=& \frac{M}{a} \frac{a}{r_H} = \frac{h}{\tilde{a}}
\ea
we integrate the angular integral
assuming the average
 magnetic field, $<B_H^2>$,
\ba
P &=&   \frac{1}{32}(\frac{a}{M})^2\frac{(2Mr_H)^2}{r_H^2}
   \frac{(1+h^2)}{h^2} <B_H^2> 2[(h+1/h)\arctan h -1]\\
  &=& \frac{1}{4}(\frac{a}{M})^2 M^2 <B_H^2> f(h)
\ea
where
\ba
f(h) &=&  \frac{(1+h^2)}{h^2} [(h+1/h)\arctan h -1]\\
     &\rightarrow& 2/3 , \,\,\,\, for \,\, h \rightarrow 0,\\
     &=& \pi -2 , \,\, for\,\, h= 1(\, extreme \, \, rotation, \, a=M)
\ea    

Now inserting G and c properly, we get
\ba
P &=& \frac{1}{4}(\frac{ac}{GM})^2 M^2 <B_H^2> f(h)\frac{G^2}{c^3}
\label{power-}\\
&=& \frac{1}{4}\tilde{a}^2 M^2 <B_H^2> f(h)\frac{G^2}{c^3}
\label{power0}\\
&=& \frac{1}{4c}\frac{J^2}{M^4} M^2 <B_H^2> f(h)\label{power}
\ea
Numerically,
\ba
\frac{G^2}{c^3} M_{\odot}^2 (10^{15}gauss)^2 &=& (2 \times 10^{33}
\times 10^{15})^2 \times \frac{(6.7 \times 10^{-8})^2}
{(3 \times 10^{10})^3}\\
  &=& 6.7 \times 10^{50} \, erg/s
\ea
Then the total power of Poynting flux is
\be
P = 1.7\times 10^{50}  (\frac{ac}{GM})^2 (\frac{M}{M_{\odot}})^2
    (\frac{<B_H>}{10^{15}gauss})^2 f(h) \, erg/s\label{powerm}
\ee
 
One of the frequently-quoted numbers for the total power of BZ process
 from ref.\cite{TPM}  is
\be
P \sim   10^{49}  (\frac{ac}{GM})^2 (\frac{M}{M_{\odot}})^2
    (\frac{<B_H>}{10^{15}gauss})^2 \, erg/s\label{kth}
\ee
which is one order of magnitude lower than eq.(\ref{power-})
\footnote{ Recently Popham
et al. \cite{popham} use
\be
P \sim   10^{50}  (\frac{ac}{GM})^2
    (\frac{<B_H>}{10^{15}gauss})^2 \, erg/s
\ee
for $M=3M_{\odot}$ hence equivalent to eq.(\ref{kth}).}.
In ref\cite{TPM}, the angular integration of eq.(\ref{ppsih}) is approximated
by replacing
\ba
\tilde{\omega}^2 &\rightarrow& r_H^2/2\\
\Delta\Psi &\rightarrow& \pi r_H^2 B_H
\ea   
\ba
P &\sim& \frac{1}{128}(\frac{ac}{GM})^2 M^2 <B_H^2> g^2(a/M)\frac{G^2}{c^3}
\label{powert}\\
g^2(ac/GM) &=& 1+(1-(ac/GM)^2)^{1/2}\\
         &\rightarrow& 1 \,\,\, for \,\, a \, \rightarrow \, 0\\
         &=& 2, \,\, for \, a = M(\, extreme \, rotation)
\ea
Then
\ba
\frac{P(eq.\ref{powert})}{P(eq.\ref{power-})} &=& \frac{g^2(ac/GM)}{32f(h)}\\
   &\rightarrow& 3/64 \,\, , for \, a \, \rightarrow \, 0\\
   &\sim& 5/56 \,\, , for \, a = M(\, extreme \, rotation)
\ea
which explains the difference of an order-of-magnitude.

\end{document}